\documentclass[aip, amssymb, amsmath, groupedaddress, reprint]{revtex4-1}

\draft

\usepackage{graphicx}
\usepackage[colorlinks=true,linkcolor=blue]{hyperref}
\usepackage{soul,xcolor}
\setstcolor{red}

\begin{document}

\title{Electromagnetic Induction Imaging with a Radio-Frequency Atomic Magnetometer}

\author{Cameron Deans}
\author{Luca Marmugi}
\email{l.marmugi@ucl.ac.uk}
\author{Sarah Hussain}
\author{Ferruccio Renzoni}
\affiliation{Department of Physics and Astronomy, University College London, Gower Street, London WC1E 6BT, United Kingdom}

\date{\today}

\begin{abstract}
We report on a compact, tunable and scalable to large arrays imaging device, based on a radio-frequency optically-pumped atomic magnetometer operating in magnetic induction tomography modality. Imaging of conductive objects is performed at room temperature, in an unshielded environment and without background subtraction. Conductivity maps of target objects exhibit not only excellent performance in terms of shape reconstruction, but also demonstrate detection of sub-millimetric cracks and penetration of conductive barriers. The results presented here demonstrate the potential of a future generation of imaging instruments, which combine magnetic induction tomography and the un-matched performance of atomic magnetometers.
\end{abstract}

\pacs{}% insert suggested PACS numbers in braces on next line

\maketitle

Imaging the properties of an object is a fundamental asset in many fields, ranging from materials' evaluation to security, and bio-medicine. Although a number of imaging techniques and devices have been developed, there are no universal imaging systems. Different systems create images based on different properties of the object of interest which, combined with practical issues such as cost-effectiveness or invasiveness, can narrow the range of possible applications.

Electromagnetic induction imaging, often referred to as Magnetic Induction Tomography (MIT)\cite{griffiths2001}, due to the possibility of producing tomographic images, is one of the best candidates for a tunable, cost-effective and non-invasive imaging tool. Indeed, its suitability to many different applications is being investigated \cite{gmrmit, zolgharni2010, darrer2015, ma2015, darrer2015aip}. MIT non-invasively maps the conductivity, permeability and permittivity of an object of interest, by measuring the ``secondary field'' ($\mathbf{B_{EC}}$) produced by eddy currents induced by an applied oscillating magnetic field. The system relies on the inductive coupling between this AC magnetic field, the ``primary field'', with the sample under investigation.

In order to overcome the limitations of current conventional MIT instrumentation, in terms of sensitivity and  bandwidth, a proof-of-concept was recently realized with a self-oscillating Optical Atomic Magnetometer (OAM)\cite{wickenbrock2014}. However, the relative complexity of the apparatus, its low suitability for scalability and its reduced flexibility in terms of working frequency, make the instrument not well-suited for practical applications.

Here, we present an alternative approach for optical MIT imaging, based on a Radio-Frequency optically pumped OAM (RF OAM)\cite{rfoam1, rfoam2, rfoam3}. This combines the advantages in terms of sensitivity of OAMs, up to orders of magnitudes larger than that of a pick-up coil of the same volume\cite{savukov2007}, with the simplicity, robustness and scalability of RF OAMs. 

We report on the implementation of an electromagnetic induction imaging system based on a RF OAM and on the demonstration of OAM-based electromagnetic imaging in conditions similar to those of materials' non-destructive evaluation and security screening. In particular, we demonstrate imaging of conductive objects, crack detection and conductive barrier penetration at room temperature, in an unshielded environment.

\begin{figure*}[htbp]
\includegraphics[width=\linewidth]{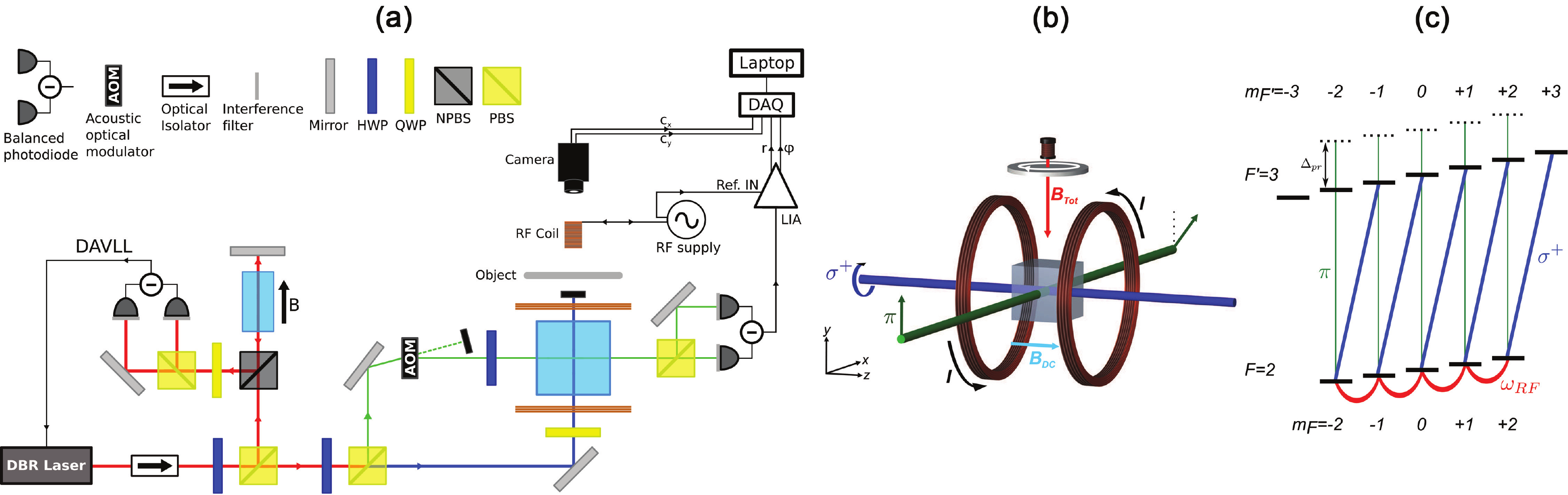}
\caption{Compact RF OAM for MIT. Pump (exactly resonant, $\sigma^{+}$ polarized, propagating along $\hat{z}$, 7.9 mW/cm$^{2}$) and probe ($\Delta_{pr}$=+425 MHz, $\pi$ polarized along $\hat{y}$, propagating along $\hat{x}$, 0.9 mW/cm$^{2}$) are tuned to the ${}^{87}$Rb $F=2\rightarrow F'=3$ hyperfine transition. They are crossed at the center of a 25 mm cubic glass cell containing 20 Torr of N$_{2}$ as buffer gas and a naturally occurring Rb vapor, kept at room temperature (298 K). A uniform DC magnetic field ($\mathbf{B_{DC}}=B_{DC}\hat{z}$) allows optical pumping to the $|F=2, m_{F}=+2\rangle$ Zeeman state and thus the spin-polarization of ${}^{87}$Rb along $\hat{z}$. An AC magnetic field  ($\mathbf{B_{RF}}$) coherently drives the ground-state polarization. Unlike in conventional RF OAMs, $\mathbf{B_{RF}}$ also acts as the primary field for MIT, thus generating eddy currents in the object of interest. Under the influence of the total magnetic field $\mathbf{B_{Tot}}=\mathbf{B_{RF}}+\mathbf{B_{EC}}$, the polarization of the probe beam is periodically rotated and measured by projecting the resulting polarization onto $\hat{z}$ and $\hat{y}$. In order to obtain position-resolved measurements, a CCD camera detects the position of the object, while a laptop records the averages ($10^{3}$ samples/point). \textbf{(a)} Experimental setup. DBR: distributed Bragg-reflector; DAVLL: dichroic atomic vapor laser lock; HWP: $\lambda/2$ waveplate; QWP: $\lambda/4$ waveplate; NPBS: non-polarizing beam-splitter; PBS: polarizing beam-splitter; LIA: lock-in amplifier; DAQ: data acquisition board. \textbf{(b)} Detail (to scale) of the MIT OAM unit, with an $Al$ disk as an example of a typical target. Eddy currents are depicted as a white loop. \textbf{(c)} ${}^{87}$Rb $F=2 \rightarrow F'=3$ transitions involved in the OAM operation. The $\pi$-polarized probe beam does not match the center of the $|F'=3, m_{F'}\rangle$ levels because of its detuning. \label{fig:setup}}
\end{figure*}

The apparatus is sketched in Fig.~\ref{fig:setup}(a). The fundamental sensing unit (Fig.~\ref{fig:setup}(b)) is an ${}^{87}$Rb vapor, spin-polarized by optical pumping via a $\sigma^{+}$ laser beam tuned to the $D_{2}$ line $F=2\rightarrow F^{'}=3$ transition (Fig.~\ref{fig:setup}(c)). Tuning is stabilized by means of an independent Dichroic Atomic Vapor Laser Lock (DAVLL). As in conventional RF OAMs, a perpendicular AC magnetic field ($\mathbf{B_{RF}}$) excites spin-coherences and produces a transverse atomic polarization. Under the action of a magnetic field to be measured, the Faraday effect rotates the plane of polarization of a $\pi$-polarized probe beam, providing information about the Larmor precession and, therefore, about the magnitude of the total magnetic field\cite{budkerbook}.  In the present work, the probe beam is detuned by +425 MHz  by a double-pass acousto-optic modulator. The effects of Faraday rotation are detected by a polarimeter, constituted of a balanced photodiode and a polarizing beam splitter. Typically, the OAM's resonance has a FWHM$\sim2\times10^{-7}$ T.

In order to reduce the system's footprint and complexity, a single ferrite-core coil (7.8 mm $\times$ 9.5 mm, L=680 $\mu$H at 1 kHz) generates both $\mathbf{B_{RF}}$ and the MIT primary field. A dual-phase lock-in amplifier, referenced to the RF supply driving the coil, allows the measurement of the amplitude (``radius'', $R$) and the phase lag (``phase'', $\phi$)  of the magnetic field signal. This includes the eddy currents' contribution  which is directly related to the dielectric properties of the object of interest\cite{griffiths1999}.

The frequency of the oscillating magnetic field ($\omega_{RF}$) is chosen on the basis of the required skin depth $\delta$. Tuning of the RF OAM is achieved by suitably adjusting the DC field used for optical pumping, $\mathbf{B_{DC}}$ (Fig.~\ref{fig:setup}(b)). Thus, the range and penetration of the system can be easily adapted to the required task and current conditions.  Position-resolved measurements are obtained by moving the samples with a translational stage.

The present set-up represents an important development in terms of a reduction of complexity with respect to the proof-of-concept\cite{wickenbrock2014}. The MIT information is now directly encoded at $\omega_{RF}$. Therefore, no down-mixing is required. Moreover, the re-designed OAM does not require phase-locked loops and feedback-controlled acousto-optic modulation for synchronous optical pumping. Furthermore, the crossed beams configuration adopted here (Fig.~\ref{fig:setup}) allows more precise control of the sensing region, thus potentially improving either the spatial resolution, the sensors' density, or both.

\begin{figure}[htbp]
\includegraphics[width=\linewidth]{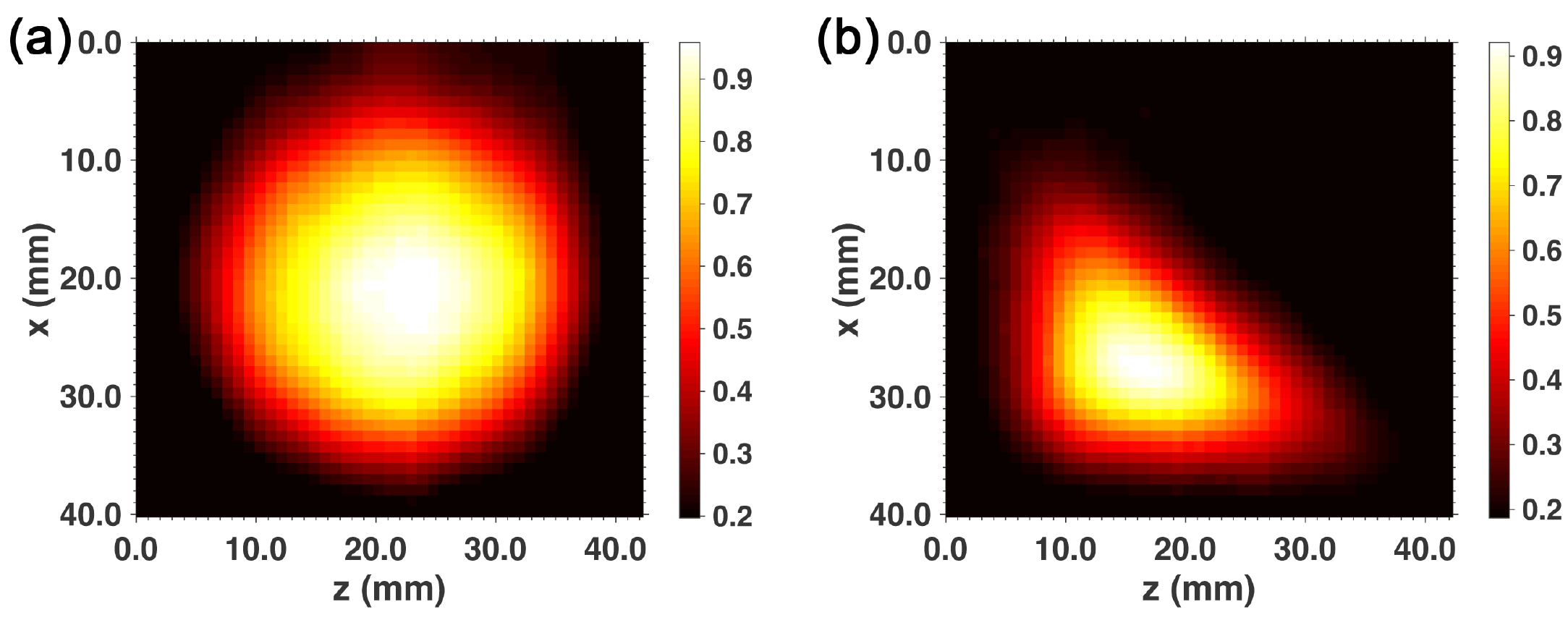}
\caption{Electromagnetic induction imaging with a RF OAM: normalized conductivity maps of Al objects at 1 kHz. \textbf{(a)} $R$ map of an Al disk, diameter 37 mm (2 mm thick). $C\approx 3.5$, as defined in Eq.~\ref{eqn:c}.  \textbf{(b)} $\phi$ map of a 31 mm $\times$ 38 mm $\times$ 50 mm Al triangle (3 mm thick).  $C\approx 3.9$. Maximum phase change is $\Delta \phi_{max}=41.3^{\circ}$, the color bar maps an interval $\Delta\phi_{col}=38.4^{\circ}$.\label{fig:imaging}}
\end{figure}

The instrument's imaging capabilities at room temperature in an unshielded environment are demonstrated by imaging several conductive objects, as shown, for example, in Fig.~\ref{fig:imaging}. We did not observe any noticeable effects during the MIT operation due to random varying stray magnetic fields. In the case of metallic and non-magnetic objects, conductivity is the largest contributor to the secondary field. The maps shown in this work can therefore be considered exclusively as plots of conductivity.

In Fig.~\ref{fig:imaging}, $\omega_{RF}/2\pi$=1 kHz, with the driving coil centered on the sensor, $63~mm$ above the pump and probe beams. The local maximum field at the object's level is estimated, by means of finite-element simulations,  to be of the order of $2\times10^{-3}$ T, and of $6\times10^{-5}$ T at the sensor's level, when the object is not present. The peak of eddy currents' surface density is estimated $\sim10^{6}$ A/m$^{2}$ in the case of Fig.~\ref{fig:imaging}(a).

The objects of interest, a disk and a triangle of Al, of 2 mm and 3 mm thickness respectively, are moved with a translational stage with respect to the RF OAM. At each position, 10$^{3}$ averages are performed. Matrices of position-dependent data $D=\{d_{zx}\}$ are acquired by a laptop, displayed in real time and recorded. Contrary to the previous implementation\cite{wickenbrock2014}, no background subtraction is required. In other words, only one image is required for reproducing objects such as those in Fig.~\ref{fig:imaging}. This is of great importance in view of practical applications.

Each matrix $D$ is normalized and  then filtered with a nearest-neighbor averaging filter, with 2 pxl radius ($d_{zx}^{(norm)}$) . This takes into account possible fluctuations in positioning and eases shapes' recognition. Data are then plotted in color-coded 2D plots. The dark level for color coding is taken as the average of the normalized values plus their standard deviation along a column $\tilde{x}$, where the object is not present: $dark \equiv \langle d^{(norm)}_{z\tilde{x}}+stdev^{(norm)}_{z\tilde{x}} \rangle$. In this way, with a confidence of 64$\%$, each value larger than $dark$ is a contribution generated by the object and not by fluctuations in the background, which are still present.  This simple criterion can be adapted in view of the desired application. It is also partially responsible for a possible lack of ``sharpness'' at the boundaries of the objects. Suitable edge detection algorithms could limit its impact. It is noteworthy that this procedure does not  modify data, only their \textit{display}. 

Finally, as a figure of merit for direct comparison of different images, we introduce the contrast $C$:

\begin{equation}
C=\dfrac{max(d_{zx}^{(norm)})- dark }{dark }~. \label{eqn:c}
\end{equation}

\begin{figure}[htbp]
\includegraphics[width=\linewidth]{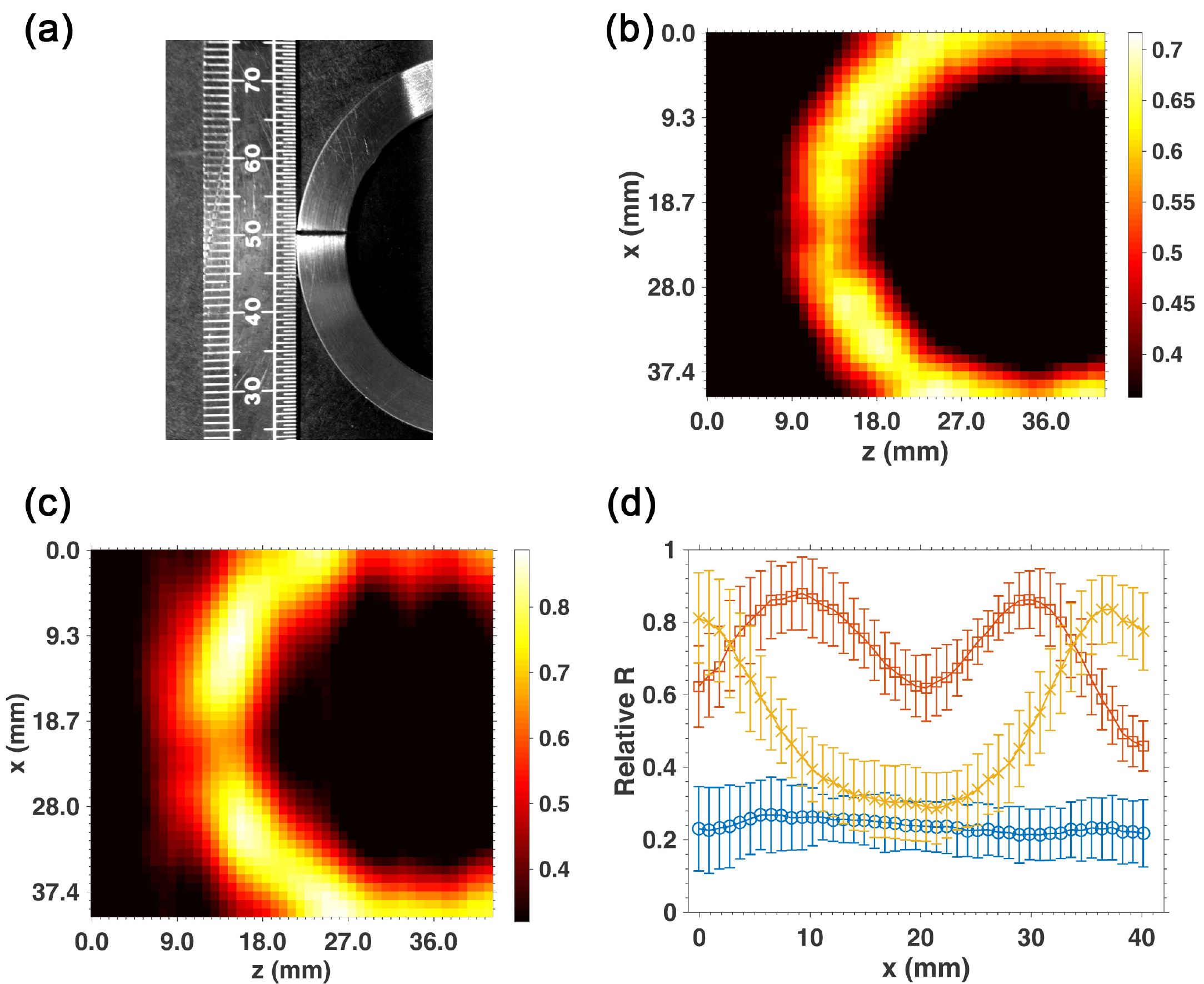}
\caption{Crack detection: normalized conductivity maps of an Al ring with a sub-mm crack at 10 kHz ($\delta_{Al}\approx 0.82$ mm). \textbf{(a)} Photo of the sample: thickness is $2~mm$, the radial extension is $6~mm$. The width of the crack is $<1$ mm. \textbf{(b)} $\phi$ map; $C\approx 1.1$.  In this case, $\Delta \phi_{max}=9.9^{\circ}$, $\Delta \phi_{col}=6.0^{\circ}$. \textbf{(c)} $R$ map; $C\approx 1.8$. \textbf{(d)} Cross-sections of the $R$ map at $z=3.2$ mm (blue circles), $z=15.2$ mm (red squares) and $z=22.4$ mm (yellow crosses). Error bars are standard deviations automatically computed after 10$^{3}$ averages. \label{fig:crack}}
\end{figure}

Non-Destructive Evaluation (NDE) is one of the main applications of electromagnetic induction imaging. In order to demonstrate the suitability of our instrument for such a task, a sub-mm crack is created in an Al ring. As shown in  Fig.~\ref{fig:crack}(a), the cut goes through the full radial extension (6 mm) and the full thickness (2 mm) of the object. This causes an interruption in the flow of the eddy currents excited by the MIT primary field, and hence a decrease of the local $\mathbf{B_{EC}}$ measured by our instrument. In terms of conductivity mapping, this will create a local minimum.

\begin{figure*}[htbp]
\includegraphics[width=\linewidth]{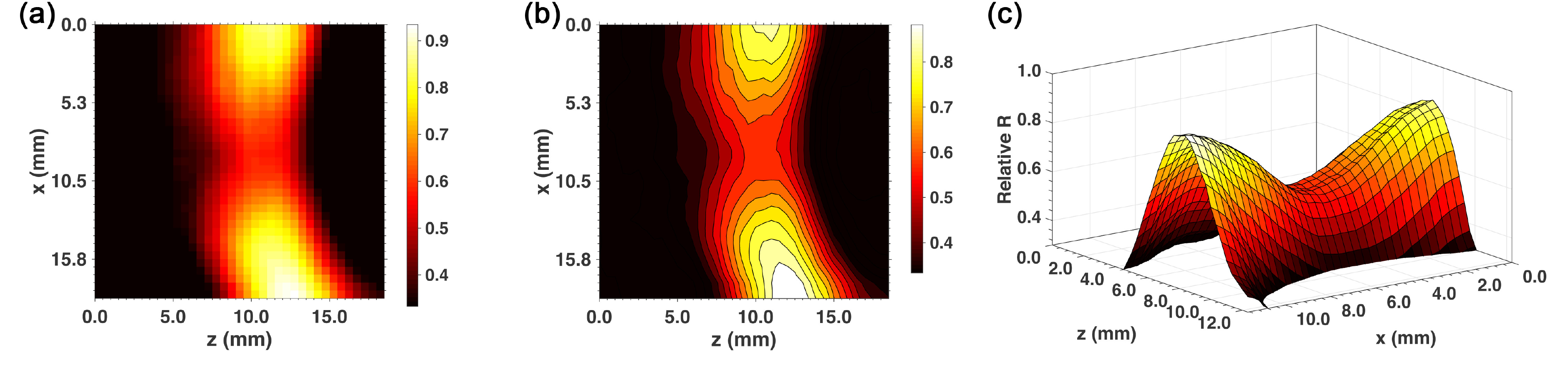}
\caption{Crack detection: high-resolution normalized conductivity map of a sub-mm crack in an Al ring at 10 kHz ($\delta_{Al}\approx 0.82$ mm). \textbf{(a)} $R$ map, $C\approx 1.8$; \textbf{(b)} Corresponding contour plot. \textbf{(c)}  Detail of a 3D surface plot ($R$) of the crack. \label{fig:crackzoom}}
\end{figure*}

Figs.~\ref{fig:crack}(b) and \ref{fig:crack}(c), which present the $\phi$ and the $R$ maps of the ring at 10 kHz, confirm this: at the position of the sub-mm crack, a decrease in the level of the signal is observed. Hence, our RF OAM-based MIT system is capable of imaging conductive objects and is also able to detect and locate tiny features, such as a structural anomalies. This is of fundamental importance for NDE, for example in industrial quality processes and structure monitoring.

Fig.~\ref{fig:crack}(d) shows a plot of normalized $R$ data, along selected columns of the radius map \ref{fig:crack}(c). The blue circles correspond to a region where the object is not present. The red squares are the cross section at the center of the crack: the trace exhibits a decrease in the conductivity at $x=$20 mm, where the fracture is located. This decrease is significant in comparison to the standard deviation bars. The yellow crosses label the cross section towards the center of the ring; as expected, a large value is read for $x\sim$0 mm and $x\sim$38 mm, whereas values consistent with the background are obtained around $x\sim$20 mm.

Incidentally, Fig.~\ref{fig:crack} also demonstrates the capability of the system to work at different frequencies, only by tuning the $\mathbf{B_{DC}}$ field.

Fig.~\ref{fig:crackzoom} further demonstrates the suitability of the system for NDE: the crack in the ring is here imaged with a higher spatial resolution. The crack is clearly visible, and is displayed in a 3D surface plot obtained from the normalized $R$ (Fig.~\ref{fig:crackzoom}(c)). As in the case of Fig.~\ref{fig:crack}, the image of the fracture is somehow smoother than expected. On the one hand, this is an obvious consequence of the nearest-neighbor averaging. On the other hand, however, this is due to the relatively large spread of the primary field at the object's surface. According to finite-element simulations, the actual distribution of $\mathbf{B_{RF}}$ on the object's plane is broader than the crack itself (FWHM$\sim$6 mm). This could be improved by using more localized driving fields, or introducing image processing which takes into account the distribution of the driving field over the spatial extension of the object.

Another fundamental capability for applications is penetrating shieldings or surrounding spurious objects. This would be of major importance for security screening  and also for the monitoring or evaluation of complex structures.

\begin{figure}[htbp]
\includegraphics[height=4.5cm]{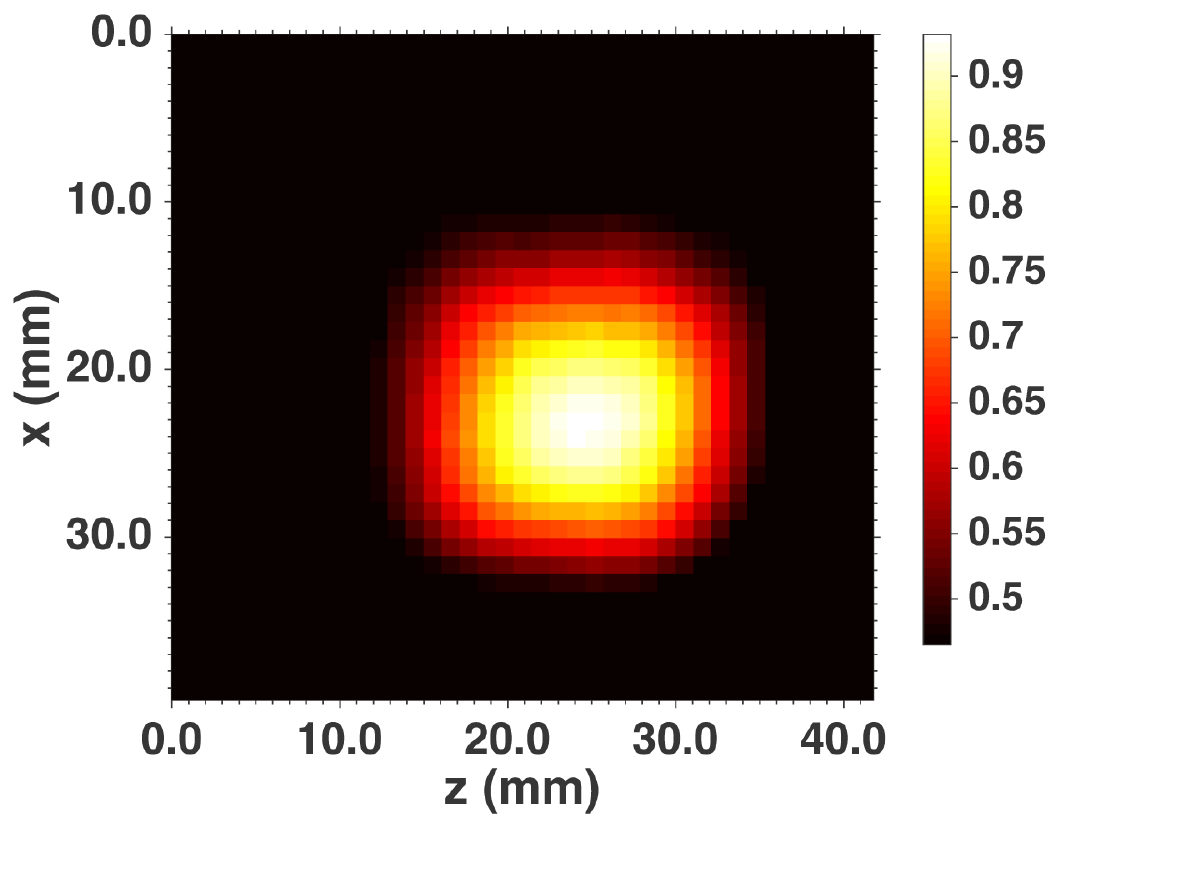}
\caption{Penetration of conductive barriers: $\phi$ map of a $Cu$ square (side-length=25 mm, thickness=1 mm) at 500 Hz through an Al barrier 1 mm thick ($\delta_{Al}\approx 3.7$ mm, $\sigma_{Al}\approx 0.38 \times10^{8}$ S/m; $\delta_{Cu}\approx2.9$ mm, $\sigma_{Cu}\approx 0.60 \times 10^{8}$ S/m). $C\approx 1.1$. Here, $\Delta \phi_{max}=33.4^{\circ}$, $\Delta \phi_{col}=25.3^{\circ}$. The object of interest and the barrier are in electrical contact. \label{fig:penetration}}
\end{figure}

The capability of our device in shield penetration is explored by considering a Cu square (side-length 25 mm, thickness 1 mm) concealed behind an Al barrier of thickness 1 mm and extended all over the sensing region ($>100$ mm).  The objects are in electrical contact: eddy currents can, in principle, flow from one surface to the other. The driving coil is lifted by 3 mm, thus leading to a center-to-center separation with respect to the sensor of 66 mm.

For similar configurations, it was found that the main problems for imaging are i) to penetrate the Al screen and ii) to eliminate its spurious contribution. The first  issue can be solved by choosing a suitable frequency which allows a skin depth larger than the thickness of the screen. In this case $\omega_{RF}/2\pi = 500$ Hz, which ensures penetration of the Al barrier.

The second problem has been solved by subtracting images obtained at different frequencies\cite{darrer2015, darrer2015aip}. However, this may be impractical for certain applications as it would require different measurements in stable conditions. Here, thanks to our RF OAM, the Cu square concealed by an Al barrier is imaged with a single measurement at 500 Hz, without the need of any background subtraction, or image reconstruction (Fig.~\ref{fig:penetration}).

The agreement with the position, the shape and the size of the object is satisfactory. Nevertheless, the edges of the square appear particularly blurred. This may be ascribed to edge effects at the boundaries between the object of interest and the screen, where further noise due to mutual induction and flow of eddy currents is expected. This is reflected in the fact that,  the $\phi$ map provides better results than the $R$ map, thanks to the different conductivities between the Al shield and the Cu square. This produces a larger phase lag in the case of the more conductive copper.

In conclusion, a compact, robust and scalable imaging instrument based on RF OAM operating in MIT modality is realized and tested in view of different applications. Sensitivity and flexibility of the system are provided by the RF OAM, which can be easily tuned in order to match the required working conditions. Imaging of conductive objects is demonstrated with such a device and without any image reconstruction or background subtraction. With respect to the proof-of-concept of MIT with OAMs\cite{wickenbrock2014}, the current system allows for improved performances in terms of spatial resolution, while reducing the complexity and the cost of the apparatus. The present work, therefore, demonstrates practical applications of MIT with RF OAMs in unscreened everyday conditions, from  security screening to non-destructive evaluation.

\begin{acknowledgments}
This work was supported by a Marie Curie International Research Staff Exchange Scheme Fellowship ``COSMA'' (PIRSES-GA-2012-295264). C.~D.~is supported by the EPSRC Centre for Doctoral Training in Delivering Quantum Technologies. L.~M.~acknowledges the support by Innovate UK within the project ``AMMIT'' (Project No. 131885). S.~H.~is supported by DSTL - Defence and Security PhD - Sensing and Navigation using Quantum 2.0 technology.
\end{acknowledgments}

\bibliographystyle{aipnum4-1}
\bibliography{apl}

\end{document}